\documentclass[preprint,12pt]{elsarticle}

\usepackage{amssymb}
\usepackage{amsmath,amsthm}
\newtheorem{theorem}{Theorem}
\newtheorem{lemma}{Lemma}
\usepackage{float}
\usepackage{graphicx}
\usepackage[T1]{fontenc}

\theoremstyle{remark}
\newtheorem{example}{Example}[section]

\journal{None}

\begin{document}

\begin{frontmatter}



\title{Strategy Evolution in the Adoption of Conservation Tillage Technology under Time Preference Heterogeneity and Lemon Market: Insights from Evolutionary Dynamics}


\author[inst1]{Dingyi Wang}

\affiliation[inst1]{organization={college of economics \& management, Northwest A\&F University},%Department and Organization
            addressline={22 Xinong Road}, 
            city={Yangling},
            postcode={712100}, 
            state={Shaanxi},
            country={China}}
            
\affiliation[inst2]{organization={Department of Mathematics and Computer Technology, Guilin Normal university},%Department and Organization
            addressline={9 Feihu Road}, 
            city={Guilin},
            postcode={541000}, 
            state={Guangxi},
            country={China}}

\author[inst2]{Ruqiang Guo\corref{cor}}
\cortext[cor]{Corresponding author}
\ead{guoruqiang227@nwafu.edu.cn}

\author[inst1]{Qian Lu\corref{cor}}
\ead{xnluqian@126.com}

\iffalse
\author[inst1]{Ruqiang Guo}

\author[inst1]{Liang Zhang\corref{cor}} 
\cortext[cor]{Corresponding author}
\ead{zhanglsd@126.com}
\fi

\begin{abstract}

The promotion of Conservation Tillage Technology (CTT) is critical for mitigating global soil degradation, yet their actual adoption rates remain substantially lower than anticipated targets. Existing research predominantly focuses on static factor analyses, failing to adequately capture the dynamic evolutionary mechanisms of farmer strategic interactions and the impacts of information asymmetries in agricultural markets. This study constructs an evolutionary game model integrating heterogeneous time preferences and lemon market effects to reveal the dynamic equilibrium of technology adoption within farmer groups operating under bounded rationality. Key findings indicate that farmers with high time preferences significantly impede CTT adoption due to the excessive discounting of long-term benefits. Furthermore, the lemon market effect dictates the system's equilibrium states: 1) When the lemon market benefit ($P$) exceeds the lemon market loss ($Q$) ($P > Q$), stable tripartite coexistence of adoption strategies emerges; 2) When $P < Q$, the system evolves unpredictably, exhibiting dynamics characterized by heteroclinic cycles; 3) At the critical threshold $P = Q$, the system transforms into a conservative Hamiltonian system, yielding stable periodic oscillation solutions. Based on these insights, policy recommendations are proposed: implementing ecological certification schemes to mitigate information asymmetry, offering subsidies and insurance to reduce adoption risks, and utilizing environmental taxes to internalize the negative externalities associated with conventional tillage. This research not only provides a dynamic analytical paradigm for the diffusion of green agricultural technologies but also furnishes a theoretical foundation for designing sustainable agricultural policies in developing countries.

\end{abstract}

\iffalse
\begin{graphicalabstract}
\end{graphicalabstract}

\begin{highlights}
\item Research highlight 1
\item Research highlight 2
\end{highlights}
\fi


\end{frontmatter}


\section{Introduction}
\label{Intr}

In recent years, the issue of low adoption rates for conservation tillage technology has been one of the core challenges in the fields of agricultural sustainable development and climate change adaptation\citep{qiu2020risk}. Excessive exploitation and utilization of water and soil resources by humans have intensified global soil erosion, posing a severe threat to global food security. According to statistics, the global area affected by soil erosion has reached $14.3$ million $km^2$ \citep{STXB202119010}, accounting for approximately $9.6$ percent of the Earth's total land surface area\citep{XBSZ202404025}. Conservation Tillage Technology (CTT), recognized for its significant ecological benefits—including reducing soil erosion, enhancing soil organic matter, improving water retention capacity, and contributing to carbon sequestration and emission reduction—has emerged as an effective approach to addressing global challenges related to food security and ecological environmental protection\citep{WOS:000078485900002}. However, globally, and particularly in developing countries, a substantial gap persists between the actual adoption rate of CTT and the targeted goals\citep{WOS:000277084100015, WOS:000785604800001}. One study revealed that in 2019, the adoption of CTT accounted for less than $6.4\%$ of China's cultivated land area\citep{2025Collaboration}. significantly below the global average. Therefore, elucidating the root causes hindering the widespread adoption of CTT by farmers and understanding their adoption decision-making mechanisms have become critical issues that urgently need resolution to effectively promote the diffusion of this technology.

Existing research on CTT primarily revolves around two main themes. The first theme focuses on identifying the influencing factors of CTT adoption. This line of research extensively investigates the key drivers and barriers affecting farmers' decisions to adopt CTT. Findings indicate that a combination of farmer characteristics (such as gender, age, education level, and off-farm employment status)\citep{2011Vitale, guo2022study}, household endowments (including income level, farm size, and cropping structure)\citep{qu2021factors, mccord2015crop}, and environmental factors (such as extension systems, technical training, rainfall and pest and disease shocks, and neighbor demonstration effects)\citep{grabowski2014resource, wang2023can, mu2024they} significantly influence farmers' CTT decisions. The second theme centers on evaluating the impacts of CTT adoption. This body of research primarily assesses the benefits arising from the implementation of CTT. Studies demonstrate that adopting CTT enables farmers to achieve economic benefits, such as increased crop yields and enhanced production profitability\citep{gathala2015conservation}. Concurrently, it delivers substantial ecological benefits, including soil erosion control, improved water use efficiency, enhanced farmland quality, strengthened soil carbon sequestration capacity, and emission reduction\citep{qin2024effects}. The aforementioned research has laid a solid foundation for understanding the adoption mechanisms and effects of CTT. However, from a methodological perspective, a crucial dynamic viewpoint—the evolutionary game perspective—remains notably underexplored within the field of CTT research.

In recent years, the evolutionary game perspective has assumed increasing significance in research on agricultural technology diffusion. Existing literature based on evolutionary game frameworks has explored the promotion and application of China's agricultural Internet of Things technology Evolution Dynamics of Agricultural Internet of Things Technology Promotion and Adoption in China\citep{2020Evolution}, as well as the impact of demand information sharing on the adoption of organic agriculture\citep{yu2021impact}. However, research specifically focused on the adoption of CTT currently lacks systematic investigation from a dynamic evolutionary game perspective. Existing static models struggle to capture the dynamic learning and behavioral adjustment processes of farmer groups under strategic interactions\citep{conley2010learning}. Moreover, the existing literature rarely considers the influence of heterogeneity in farmers' time preferences and the lemon market effect on technology diffusion. A recent study by Esau Simutowe et al. (2024) demonstrates that farmers with a preference for immediate consumption (i.e., higher discount rates) exhibit impatience towards future investments, which discourages the adoption of CTT\citep{simutowe2024risk}. Furthermore, the lemon market effect, arising from information asymmetry, also hinders technology adoption\citep{kouser2020evaluating}. This combination of overlooked factors has resulted in notable limitations in the current understanding of CTT adoption.

In light of these limitations, this study constructs an evolutionary game model grounded in evolutionary game theory, systematically incorporating heterogeneity in time preferences, technology externalities, and information asymmetry in agricultural product markets into the analytical framework to investigate in depth the evolutionary mechanisms underlying farmers' adoption behavior of CTT and identify its stable equilibrium conditions. The model captures the dynamic process whereby a population of boundedly rational farmers, seeking to maximize their individual utility, adaptively adjust their level of adoption of CTT—a continuous strategy—through learning. Specifically, it addresses three key research questions: (1) How does heterogeneity in time preferences influence individual farmers' adoption decisions regarding CTT? (2) How does the lemon market effect impede the effective diffusion of CTT? (3) What are the key threshold conditions for achieving the ideal state of ``complete adoption" of CTT? Our findings reveal that farmers' time preferences affect their adoption level by influencing the present value of the future benefits derived from adoption; the magnitude of the lemon market effect significantly influences the stability of intra-population equilibrium points; and we demonstrate the existence of key threshold conditions necessary for achieving the ideal state of complete adoption by all farmers, providing valuable insights for designing policy interventions.

Compared to existing research, the marginal contributions of this study are twofold. First, it breaks through the limitations of traditional static analysis frameworks by introducing evolutionary game theory into the study of CTT adoption. By constructing an evolutionary game model of bounded rationality among farmer groups, this research simulates the mechanisms of learning, imitation, and behavioral adjustment during the technology adoption process. This approach captures the long-term feedback pathways between individual decision-making and group evolution, thereby providing a more explanatory and universally applicable dynamic analytical framework for analyzing the diffusion of agricultural environmental technology innovations exhibiting similar characteristics. Second, this study integrates time preference heterogeneity and the lemon market effect within a unified framework. Specifically, it incorporates heterogeneity in time preferences to quantify farmers' intertemporal trade-offs between the long-term ecological benefits and short-term learning costs of CTT under varying discount rates, revealing the nonlinear impacts of farmers' patience levels (discount rates) on adoption intensity. Furthermore, it embeds the lemon market effect to characterize the absence of quality premiums for agricultural products resulting from information asymmetry, elucidating the deep-seated obstacles that this market failure poses to CTT diffusion.

The specific content distribution is as follows: In Sect. \ref{sec:Model}, we propose a model to examine the dynamics of farmers' adoption of CTT. Subsequently, in Sect. \ref{sec:Results}, we conduct a theoretical analysis of our model and provide numerical simulations within the same section to illustrate our theoretical findings. Finally, in Sect. \ref{sec:Conclusion}, we present the concluding remarks of our study.

\section{Model and method}
\label{sec:Model}
\subsection{Game of adoption of conservation tillage technology}

This study constructs an evolutionary game model to analyze farmers' adoption strategies for CTT, framed within evolutionary game theory and integrating key factors such as time preferences, technology externalities, and the lemon market effect. The Food and Agriculture Organization classifies CTT into five main categories: No-Tillage (NT), Mulch Tillage (MT), Strip Tillage (ST), Ridge Tillage (RT), and Reduced/Minimum Tillage (RMT)\citep{agriculture1993soil}. Empirical evidence suggests that the combined application of multiple CTT practices often yields more significant synergistic benefits. Therefore, based on the actual level of adoption among farmers, we categorize the strategic choices into the following three types: Complete adoption of CTT ($C$), defined as adopting $4$ or $5$ CTT categories; Partial adoption of CTT ($P$), defined as adopting $1$ to $3$ CTT categories; and Non-adoption of CTT ($N$), defined as adopting none of the CTT categories.

Time preference, a fundamental concept in behavioral economics, describes an individual's inherent tendency to value immediate rewards more highly than delayed rewards\citep{WOS:000388383700031}. In neoclassical economics, the discount rate and personal time preference are typically modeled as a same parameter to capture the trade-off between current and future consumption within an individual’s utility function\citep{Laubach2003}. Thus, in this study, time preference is quantified using the discount rate r. The higher the degree of time preference, the larger the discount rate $r$, meaning individuals place greater value on immediate gains, while the present value of future gains diminishes accordingly\citep{WOS:000988663400001}. Farmers with high $r$ favor short-term technologies like conventional tillage, while those with low $r$ prioritize long-term gains, willingly incurring learning costs for future CTT benefits. Crucially, CTT's benefit cycle lengthens progressively with adoption intensity: strategy $C$ involves high technological integration complexity, yielding returns $R_1$ after a $2$-year cycle (discount factor: $\dfrac{1}{(1+r)^2}$) with substantial learning costs $C_1$; strategy $P$ generates $R_2$ after a $1$-year cycle (discount factor:$\dfrac{1}{1+r}$) with reduced learning costs $C_2$ ($C_1 > C_2$); strategy $N$ provides undiscounted returns $R$ without learning costs ($R_1>R_2>R$).

Externality occurs when an economic activity by a microeconomic agent affects other members of society without corresponding compensation or liability, encompassing both positive and negative forms. Positive externality denotes economic behaviors generating beneficial impacts on others or the environment, while negative externality implies detrimental consequences\citep{pajewski2020measuring}. Under strategy $C$, ecological improvements (e.g., enhanced soil and water quality) create positive externality $F_1$ benefiting adjacent farmers. Conversely, strategy $N$ induces ecological degradation (e.g., soil erosion), generating negative externality $F_2$ that harms neighbors. strategy $P$, given its intermediate technical effects, is assumed to yield neutral externality ($0$). When adjacent farmers $A$ and $B$ interact: If $A$ chooses strategy $N$ while $B$ chooses strategy $C$, A gains the positive externality $F_1$ generated by $B$ without cost, while $B$ incurs the negative externality $F_2$ caused by $A$.

Agricultural producers have complete knowledge of the hazardous chemicals used during production, whereas consumers cannot discern whether agricultural products are safe and high-quality before purchase. This information asymmetry between buyers and sellers leads to adverse selection\citep{mocan2007can}, resulting in a “lemon market”. Adopting conservation tillage technology incurs higher costs. Farmers who completely adopt this technology are squeezed out of the market due to the high prices of their premium products, thereby incurring a lemon market loss $Q$. Conversely, non-adopters sell their products at lower prices and gain a lemon market benefit $P$. These gains ($P$) and losses ($Q$) are collectively termed the lemon market effect in this study. When adjacent farmers $A$ and $B$ interact: If $A$ chooses non-adoption (Strategy $N$) and $B$ chooses complete adoption (Strategy $C$), $A$ gains the lemon market benefit $P$, while $B$ bears the lemon market loss $Q$. This outcome reflects the intensity of market failure. Parameter Constraints: $R_1, R_2, R, r, C_1, C_2, F_1, F_2, P, Q>0$; $R_1 > R_2 > R, C_1 > C_2, R_1>C_1, R_2> C_2$, and $0 < r < 1$. To enhance comprehension of the model employed in this study, the table below lists all parameters and their definitions.

%%==========================
\begin{table}[H]
    \centering 
    \caption{\label{Mp1} Meaning of each parameter in the model.}
    \begin{tabular}{cc}
    \hline
    Parameters &  Definitions   \\
    \hline
    $R_1$      & Payoffs of completely adopting conservation tillage technology         \\
    $R_2$      & Payoffs of partially adopting conservation tillage technology          \\
    $R$        & Payoffs of not adopting conservation tillage technology                \\
    $r$        & Discount rate for adopting conservation tillage technology        \\
    $C_1$      & Learning cost for completely adopting conservation tillage technology   \\
    $C_2$      & Learning cost for partially adopting conservation tillage technology    \\
    $F_1$      & Positive externality from completely adopting conservation tillage technology\\
    $F_2$      & Negative externality from not adopting conservation tillage technology  \\
    $P$        & Lemon market benefit                                                 \\
    $Q$        & Lemon market loss                                                 \\
    \hline
    \end{tabular}
\end{table}
%%==========================

The payoffs matrix is represented as:
\begin{table}[H]
    \renewcommand{\arraystretch}{1.8}
    \centering 
    \caption{\label{payoffs} Payoffs matrix}
    \begin{tabular}{c|ccc}
    \hline
                 & Strategy $C$ &  Strategy $P$ & Strategy $N$  \\
    \hline
    Strategy $C$ & $\dfrac{R_1}{(1+r)^2}-C_1+F_1$ & $\dfrac{R_1}{(1+r)^2}-C_1$ & $\dfrac{R_1}{(1+r)^2}-C_1-F_2-Q$\\

    Strategy $P$ & $\dfrac{R_2}{1+r}-C_2+F_1$ & $\dfrac{R_2}{1+r}-C_2$ & $\dfrac{R_2}{1+r}-C_2-F_2$\\

    Strategy $N$ & $R+F_1+P$ & $R$ & $R-F_2$\\
    \hline
    \end{tabular}
\end{table}

For convenience in subsequent discussions, we let $M = \dfrac{R_1}{(1+r)^2}-C_1$ and $N = \dfrac{R_2}{1+r}-C_2$. Therefore, the payoffs matrix can be written as:
$$\begin{pmatrix}
M+F_1 & M & M-F_2-Q\\
N+F_1 & N & N-F_2\\
R+F_1+P & R & R-F_2
\end{pmatrix}.$$

\subsection{Replicator equations}
To study the evolutionary dynamics of the trust game, we adopt the replicator equations\citep{taylor1978}. The proportions of complete adopters$(C)$, partial adopters$(P)$, and non-adopters$(N)$ in the population are represented by $x, y, z$ respectively. Therefore, we have $x + y + z = 1$. Consequently, the evolutionary dynamics of trust can be described by the following equations\citep{huang2018,liu2018,WANG2021123944,SHI2023137720,GUO2023114078,Szolnoki_2010, SzolnokiCyclicdominance}:
\begin{equation}
    \begin{cases}
      \dot{x}=x(f_x-\phi) \\
      \dot{y}=y(f_y-\phi) \\
      \dot{z}=z(f_z-\phi),
    \end{cases}
\end{equation}

where $f_j$ denotes the expected payoffs of strategy $j=x,y,z.$ $\phi=x f_x+y f_y+z f_z$ denotes the average payoffs of the whole population.\par
The expected payoffs of $C-$strategy can be given by
\begin{align*}
f_{x}=&x\cdot (M+F_1) +y\cdot M + z\cdot (M-F_2-Q)\\
=&M+xF_1-zF_2-zQ.
\end{align*}\par
The expected payoffs of $P-$strategy can be given by
\begin{align*}
f_{y}=&x\cdot (N+F_1) +y\cdot N + z\cdot (N-F_2)\\
=&N+xF_1-zF_2.
\end{align*}
while that of $N-$strategy is given by
\begin{align*}
f_{z}=&x\cdot (R+F_1+P) +y\cdot R + z\cdot (R-F_2)\\
=&R+xF_1-zF_2+xP.
\end{align*}\par
Due to the relation $y=1-x-z,$ the original equations can be decoupled as follows,
\begin{equation}
    \begin{cases}
      \dot{x}=x(f_x-\phi) \\
      \dot{y}=y(f_y-\phi),
    \end{cases}
\end{equation}

where $\phi=x f_x+y f_y+(1-x-y) f_z$. Therefore, the three-strategy model we investigate can be simplified into a two-dimensional dynamical system of differential equations.
\begin{equation}
\label{eqsys}
\begin{cases}
  \dot{x}=x\left[(1-x)(f_x-f_z)-y(f_y-f_z) \right]\\
  \dot{y}=y\left[(1-y)(f_y-f_z)-x(f_x-f_z) \right],
\end{cases}
\end{equation}
where
$$f_x-f_z=M-R-z Q-x P=M-R-(1-x-y) Q-x P\triangleq \varphi(x,y)$$
and
$$f_y-f_z=N-R-x P\triangleq \psi(x).$$

In this study, we investigate the evolutionary dynamics of the adoption levels of conservation tillage technology among farmers within an infinite well-mixed population, employing both theoretical analysis and numerical simulations.
\section{Results}
\label{sec:Results}
\subsection{Stability analysis of the equilibrium points in System (\ref{eqsys})}
Due to the highly non-linear mathematical expression of the payoff function in the population, the conditions of $R=N$ and $N=M$ for the existence of equilibrium points at $x=0$ and $z=0$ are too stringent and not practically meaningful. Therefore, this paper assumes $R\neq N$ and $N\neq M$, as detailed in \ref{appendix1}.\par
Based on the given assumptions, System (\ref{eqsys}) has at most five equilibrium points, as detailed in \ref{appendix1}. There are three trivial equilibrium points: Equilibrium $C$, represented as $(1,0,0)$; Equilibrium $P$, represented as $(0,1,0)$; and Equilibrium $N$, represented as $(0,0,1)$. 
When $Q<M-R<P$ or $P<M-R<Q$, there exists a unique fixed point on the $CN$ boundary, given by $(x_0, 0, 1-x_0)$, where $x_0=\frac{M-R-Q}{P-Q}$.
When $M>N>R$ and $0<Q(N-R)+P(M-N)<PQ$, there exists a unique equilibrium point within the region, given by $(x^*, y^*, z^*)$, where $x^*=\frac{N-R}{P},z^*=\frac{M-N}{Q}$ and $y^*=1-x^*-z^*$.\par
By calculating the eigenvalues of the Jacobian matrix for each equilibrium point, the local dynamics in the vicinity of these points can be determined. This is crucial for understanding and predicting the system's dynamical evolution. Next, the Jacobian matrix and its elements for the equation system (\ref{eqsys}) are presented.
\begin{equation}
J=\begin{pmatrix}
    J_{11}&J_{12}\\
    J_{21}&J_{22}
\end{pmatrix}.
\end{equation}
where
$$
\begin{aligned}
J_{11}=\frac{\partial \dot{x}}{\partial x}=&[(1-x)\varphi(x, y)-y\psi(x)]+ x\left[-\varphi(x, y)+(1-x)\varphi_x'(x, y)-y\psi'(x))\right], \\
J_{12}=\frac{\partial \dot{x}}{\partial y}=& x\left[-\psi(x)+(1-x)\varphi_y'(x, y)\right],\\
J_{21}=\frac{\partial \dot{y}}{\partial x}=& y\left[-\varphi(x, y)+(1-y)\psi'(x)-x\varphi_x'(x, y)\right],\\
J_{22}=\frac{\partial \dot{y}}{\partial y}=&[(1-y)\psi(x)-x\varphi(x, y)]+ y\left[-\psi(x)-x\varphi_y'(x, y))\right]. \\
\end{aligned}
$$\par
Now we will provide the Jacobian matrix for each equilibrium point and analyze its stability.\par
The Jacobian matrix of the system at $(x, y, z) = (1, 0, 0)$ is given by 
\begin{equation}
J(1, 0, 0) = \begin{pmatrix}
-(M-R-P) & * \\
0 & N-M
\end{pmatrix}
\end{equation} 
Therefore, when $M-R-P>0$ and $N<M$, the equilibrium point of system (\ref{eqsys}) is stable.\par
The Jacobian matrix of the system at $(x, y, z) = (0, 1,0)$ is given by 
\begin{equation}
J(0,1,0)=\begin{pmatrix}
    M-N&0\\
    *&R-N
\end{pmatrix}.
\end{equation}
When $N>R$ and $N>M$, the equilibrium point of system (\ref{eqsys}) is stable.\par
The Jacobian matrix of the system at $(x, y, z) = (0,0,1)$ is given by 
\begin{equation}
J(0,0,1)=\begin{pmatrix}
    M-R-Q&0\\
    0&N-R
\end{pmatrix}.
\end{equation}
When $M-R<Q$ and $N<R$, the equilibrium point of system (\ref{eqsys}) is stable.\par

When $Q<M-R<P$ or $P<M-R<Q$, for the boundary equilibrium point $(x, y, z) = (x_0, 0, 1-x_0)$, the Jacobian matrix is 
\begin{equation}
J(x_0, 0, 1-x_0)=\begin{pmatrix}
    x_0(1-x_0)(Q-P)&0\\
    *&N-R-x_0P
\end{pmatrix}.
\end{equation}\par
We can determine that the equilibrium point is unstable when $Q>P$, and when $Q<P$ and $Q(N-R)+P(M-N)>PQ$, the equilibrium point is stable.\par
When $M>N>R$ and $0<Q(N-R)+P(M-N)<PQ$, for the internal equilibrium point $(x, y, z) = (x^*, y^*, z^*)$, the Jacobian matrix is 
\begin{equation}
J(x^*, y^*, z^*)=\begin{pmatrix}
    x^*(1-x^*)(Q-P)+x^*y^*P & x^*(1-x^*)Q\\
    -y^*(1-y^*)P-x^*y^*(Q-P)&-x^*y^*Q
\end{pmatrix}.
\end{equation}\par
According to the sizes of P and Q, there are three different dynamics behaviors:
\begin{enumerate}
    \renewcommand{\labelenumi}{($\theenumi$)}
    \item When $Q<P$, the equilibrium point $(x^*, y^*, z^*)$ is stable.
    \item When $Q>P$, the equilibrium point $(x^*, y^*, z^*)$ is unstable.
    \item When $Q=P$, according to the Jacobian matrix of system (\ref{eqsys}), it can be observed that $|J|>0$ and $tr(J)=0$. Therefore, there are no real eigenvalues at this equilibrium point, indicating that it is a center. However, for nonlinear systems, the linearized center may not necessarily be the center of the original system. Therefore, this paper provides Theorem \ref{thm1} to prove that this point is a center, indicating that system (\ref{eqsys}) is a conservative Hamiltonian system.
\end{enumerate}

\begin{lemma}\label{lemma1}
Let 
    \begin{equation}\label{eq1}
        \begin{cases}
          \dfrac{dx}{dt}=-y+a_{20}x^2+a_{11}xy+a_{02}y^2\\
          \dfrac{dy}{dt}=x+b_{20}x^2+b_{11}xy+b_{02}y^2,
        \end{cases}
    \end{equation}
and take $W_1=A\alpha-B\beta$, where $A=a_{20}+a_{02}, B=b_{20}+b_{02}$. When $A=B=0$ and $W_1 = 0$, the origin becomes the center point.
\end{lemma}
The proof of the lemma can be found in reference\citep{ma2001}.
\begin{theorem}\label{thm1}
    When $P=Q$, the system
    \begin{equation}\label{eq2}
        \begin{cases}
          \dfrac{dx}{dt}=x\left[(1-x)\varphi(y)-y\psi(x) \right]\\
          \dfrac{dy}{dt}=y\left[(1-y)\psi(x)-x\varphi(y) \right],
        \end{cases}
    \end{equation}
    is a conservative Hamiltonian system, with $\varphi(y)=M-R-zP-xP=M-R-(1-y)P$, $\psi(x)=N-R-xP$.
\end{theorem}
\begin{proof}
    Let $X=x-x^*,Y=x-x^*$, then the system (\ref{eq2}) can be written as
\begin{equation}\label{eq3}
    \begin{cases}
        \dfrac{d X}{d t}=P(X+x^*)\left[(1-x^*)Y+y^*X\right] \\
        \dfrac{d Y}{d t}=-P(Y+y^*)\left[(1-y^*)X+x^*Y\right].
    \end{cases}
\end{equation}
    The equilibrium point $(x^*, y^*,z^*)$ is translated to the origin $O$. Then a non-degenerate linear transformation is taken: 
\begin{equation*}
    \begin{cases}
    \xi=Py^*(1-y^*)x+Px^*y^*y \\
    \eta=\sqrt{\Delta}y
    \end{cases}
\end{equation*}
where $\Delta=P^2x^*y^*\sqrt{1-x^*-y^*}$. System \ref{eq3} has been transformed into standard form: 
\begin{equation}\label{eq4}
    \begin{cases}
        \dfrac{d\xi}{d t}=\sqrt{\Delta}\eta+\dfrac{\xi^2}{1-y^*}+\dfrac{P(1-2x^*-y^*)}{(1-y^*)\sqrt{\Delta}}\xi\eta -\dfrac{\eta^2}{1-y^*}\\
        \dfrac{d\eta}{d t}=-\sqrt{\Delta}\xi-\dfrac{\xi\eta}{y^*}.
    \end{cases}
\end{equation}
    For system (\ref{eq4}), $A=\frac{1}{1-y^*}-\frac{1}{1-y^*}=0$, $B=0$, and $W_1=0$, it can be deduced from Lemma \ref{lemma1} that the origin is the center, i.e., the equilibrium point $(x^*, y^*,z^*)$ is the center of the nonlinear system.
\end{proof}
The above theorem indicates that a system initialized with a point inside $S_3$ (not an equilibrium point) will form closed periodic orbits.

\subsection{Numerical simulation}
Next, we utilize EGTtools\citep{EGTtools} to conduct numerical simulations to verify the above conclusions. The three corner equilibria of the system always exist. If $M > N > R$ and $P(M-N) + Q(N-R) < PQ$, then the system will exhibit an internal equilibrium point. In addition, when $P > Q$, the internal equilibrium point of the system is stable; when $P = Q$, the internal equilibrium point is a center; when $P < Q$, the internal equilibrium point of the system is unstable. When $P < M-R < Q$ or $Q < M-R < P$, equilibrium points exist on the $CN$ boundary. When $Q < M-R < P, M > N > R$, and $P(M-N) + Q(N-R) > PQ$ or $Q < M-R < P, M > R > N$, equilibrium points on the $CN$ boundary are stable. Here, $M = \dfrac{R_1}{(1+r)^2}- C_1, N =\dfrac{R_2}{1+r} - C_2$.
For the ease of analysis, we assume $R=10, R_2=2R=20, R_1=3R=30, C_2=5, C_1=2C_2=10$. Based on these assumptions, we have $M=\dfrac{3R}{(1+r)^2}-2C_2=\dfrac{30}{(1+r)^2}-10, N=\dfrac{2R}{1+r}-C_2=\dfrac{20}{1+r}-5, R=10$. From the expressions above, it can be seen that time preference (represented by the discount rate $r$) is the most important factor affecting the payoffs of the three adoption levels. As the value of r gradually increases from $0$ to $1$, there will be respectively the maximum values of $M$, $N$, and $R$ for the payoffs of the three adoption levels. When $r\in(0, 0.162)$, M is at its maximum; when $r\in (0.162, 0.333),$ N is at its maximum; when $r\in(0.333, 1)$, $R$ is at its maximum.

\begin{figure}[H]
\centering
\includegraphics[width=1.0\textwidth]{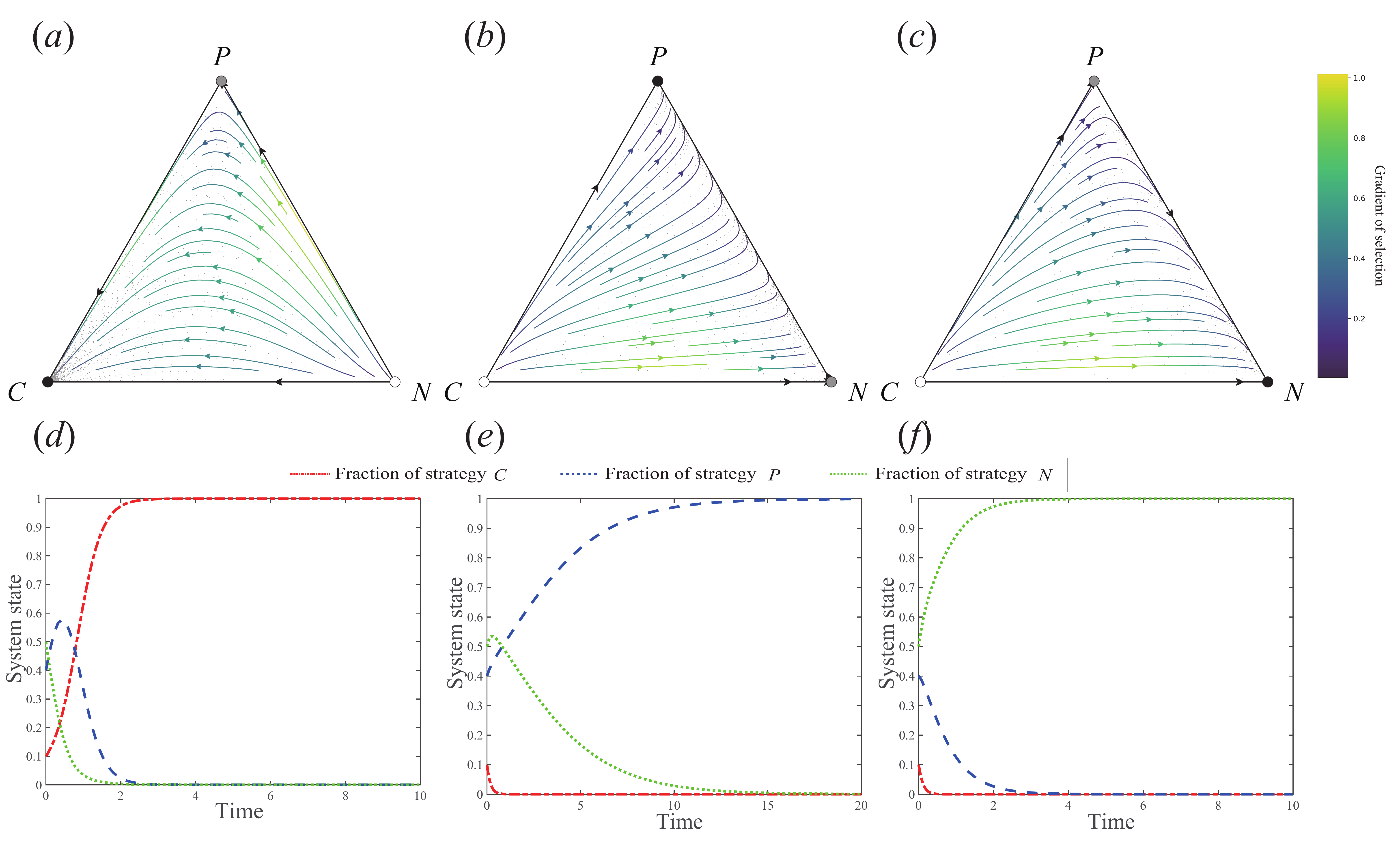}
\caption{\label{fig1} The population dynamics of different adoption levels of conservation tillage technology, namely complete adoption ($C$), partial adoption ($P$), and non-adoption ($N$), are described based on different time preferences ($r$) in $S_3$. $x, y, z$ represent the frequencies of the three strategies $C, P, N$, and $S_3 = \{(x, y, z): x, y, z \geq 0, x + y + z = 1\}$ is a triangular state space of the system. Stable and unstable equilibrium points are represented by filled circles and hollow circles, respectively. In Panel ($a$) and ($d$), $r = 0.05$, in Panel ($b$) and ($e$), $r = 0.3$, and in Panel ($c$) and ($f$), $r = 0.5$. Other parameter values include: $R = 10, C_2 = 5, F_1 = 5, F_2 = 8, Q = 6$, and $P = 5$.}
\end{figure}

\begin{figure}[H]
\centering
\includegraphics[width=1.0\textwidth]{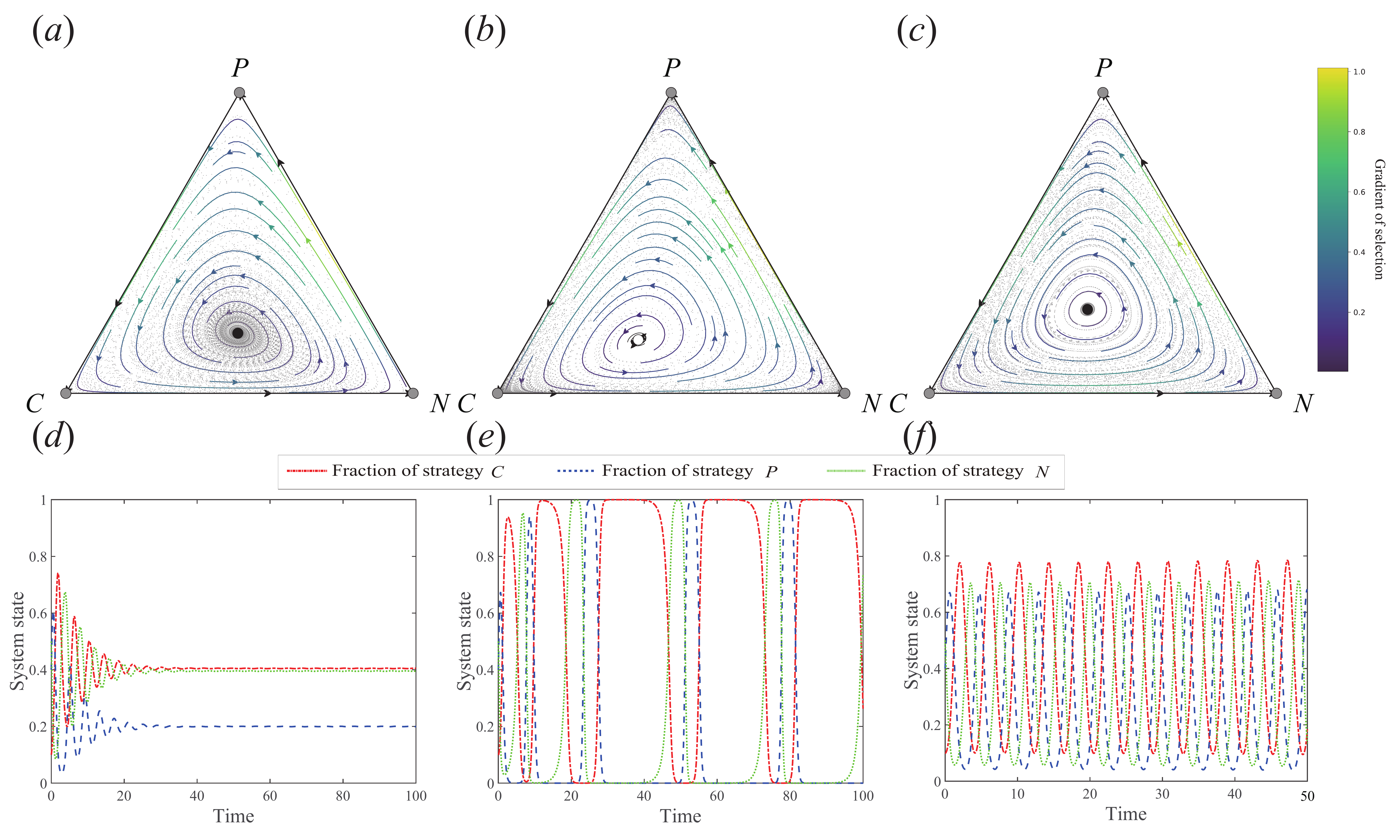}
\caption{\label{fig2} In Panel ($a$) and ($d$), $Q = 8$ and $P = 10$, in Panel ($b$) and ($e$), $Q = 10$ and $P = 8$, and in Panel ($c$) and ($f$), $Q = 10$ and $P = 10$. Other parameter values include: $R = 10, C_2 = 5, F_1 = 5, F_2 = 8$, and $r = 0.05$.}
\end{figure}

%%example 1
\begin{example}
\begin{enumerate}
    \renewcommand{\labelenumi}{$(\theenumi)$}
    \item When $M - R > \max\{Q,P\} \land M > N$, the system retains only three corner equilibria. Among these, only $(1, 0, 0)$ is asymptotically stable, while $(0, 1, 0)$ and $(0, 0, 1)$ are unstable. As shown in Fig. \ref{fig1}(a), all evolutionary trajectories converge to $(1, 0, 0)$ regardless of initial conditions, indicating complete adoption of conservation tillage by all farmers, with other strategies vanishing.
    \item If $M < N$, $R < N$, and either $M - R > Q, M - R > P$ or $M - R < Q, M - R < P$, the system still exhibits three corner equilibria. Stability shifts to $(0, 1, 0)$, which becomes the sole stable equilibrium (Fig. \ref{fig1}(b)). In other words, eventually, all farmers will partially adopt conservation tillage technology, and the other two strategies will disappear.
    \item For $M - R < \min\{Q,P\} \cap R > N$, the stable equilibrium transitions to $(0, 0, 1)$. Fig. \ref{fig1}(c) demonstrates global convergence to this state, reflecting total abandonment of conservation tillage technology. Theoretical validations are provided in \ref{appendix2} (Cases 1-4).
\end{enumerate}\par
In summary, the higher the time preference of the farmers, the lower their willingness to adopt conservation tillage technology.
\end{example}
%======================= fig 1 ==========================
%==================================================================
%%example 2
\begin{example}
Under the condition $ M - R < \min\{Q,P\} \land M > N > R $, the system exhibits four equilibria: three unstable vertex equilibria $(1,0,0)$, $(0,1,0)$, $(0,0,1)$ and one interior equilibrium. The stability of the interior equilibrium is governed by the relative magnitudes of $ P $ and $ Q $.
\begin{enumerate}
    \renewcommand{\labelenumi}{$(\theenumi)$}
    \item For $ P > Q $, the interior equilibrium satisfies $ |J| > 0 $, $ \mathrm{tr}(J) < 0 $, implying asymptotic stability. This corresponds to coexistence of all three strategies (complete/partial/non-adoption of conservation tillage) in a dynamically balanced proportion, as visualized in Fig. \ref{fig2}(a).
    \item When $ P < Q $, the interior equilibrium becomes unstable ($ |J| > 0 $, $ \mathrm{tr}(J) > 0 $), and the system forms a heteroclinic cycle connecting the vertex equilibria and boundary saddles. Fig. \ref{fig2}(b) illustrates this evolutionary indeterminacy, where no strategy achieves dominance—a phenomenon distinct from chaos despite deterministic dynamics \citep{inbook}.
        
    \item When $P=Q$, the internal equilibrium point corresponds to $|J|>0, tr(J)=0$, which means the internal equilibrium point is a center of the system. In other words, when the positive bidding effect is equal to the negative bidding effect, the system is a Hamiltonian system. As shown in Figure \ref{fig2}(c), in the phase space, the system has a family of periodic closed orbits outside the internal equilibrium point. We observe that the frequency of farmers choosing the three strategies exhibits periodic oscillation. Theorem \ref{thm1} provides the theoretical basis for the above results. 
\end{enumerate}\par
\end{example}
%======================= fig 2 ==========================
%==================================================================
%%example 3
\begin{example}

\begin{enumerate}
    \renewcommand{\labelenumi}{$(\theenumi)$}
    \item Under the regime $ P < M - R < Q \land M > N > R \land P(M - N) + Q(N - R) > PQ $, the system contains four equilibria: three vertex equilibria $(1,0,0)$, $(0,1,0)$, $(0,0,1)$ and one boundary equilibrium on edge $CN$. The equilibrium point $(1, 0, 0)$ is a stable corner equilibrium point, while the other three equilibrium points are unstable. Absence of stable interior equilibria ensures global convergence to completely conservation tillage adoption regardless of the initial frequency values of the farmers (Fig. \ref{fig3}(a)). Theoretical analysis in  \ref{appendix2} (Case 6) confirms this monostable regime.
    \item When $ P < M - R < Q \land M > N \land N < R $, the system exhibits bistability with two stable equilibrium point $(1,0,0)$ and $(0,0,1)$, coexisting with unstable equilibrium point $(0,1,0)$ and a boundary equilibrium on the $CN$ boundary. Phase trajectories bifurcate between the basins of attraction (Fig. \ref{fig3}(b)), demonstrating initial-condition-dependent outcomes. This bistable regime is formally derived in \ref{appendix2} (Case 8).

    \item For either $ (P < M - R < Q \land M < N \land N > R) $ or $ (Q < M - R < P \land N > M > R) $, the system converges exclusively to partial conservation tillage adoption. The system has one stable equilibrium point $(0,1,0)$, with three unstable equilibrium points $(1,0,0)$, $(0,0,1)$, and a boundary equilibrium on edge $CN$ (Fig. \ref{fig3}(c)). For detailed theoretical proofs, please refer to Case 9 and Case 10 in \ref{appendix2}.
\end{enumerate}\par
\end{example}
%======================= fig 3 ==========================
\begin{figure}[H]
\centering
\includegraphics[width=1.0\textwidth]{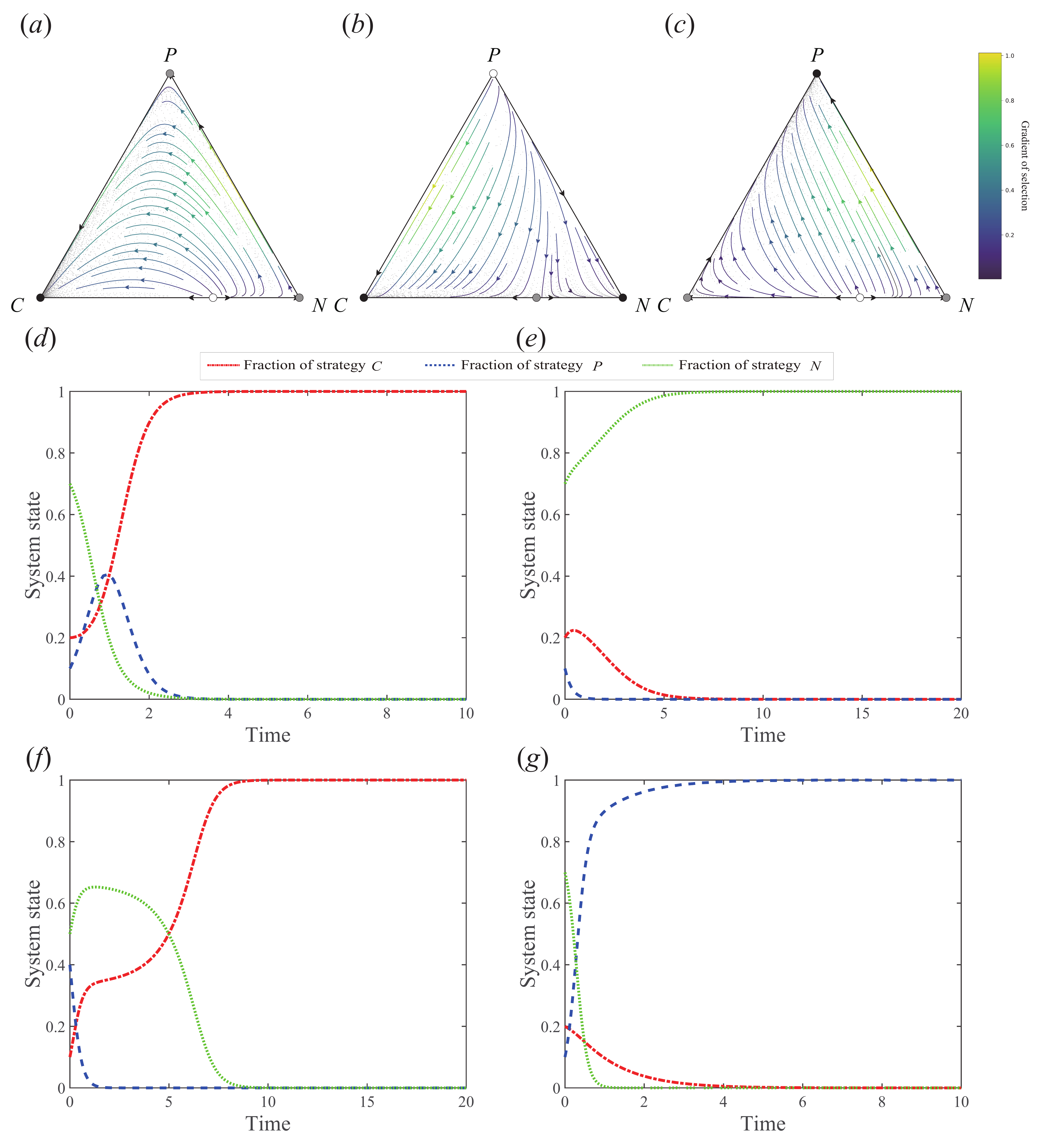}
\caption{\label{fig3} In Panel ($a$) and ($d$), $M = 17$ and $N = 14$, in Panel ($b$), ($e$) and ($f$), $M = 17$ and $N = 8$, and in Panel ($c$) and ($g$), $M = 17$ and $N = 18$. Other parameter values include: $R = 10, C_2 = 5, F_1 = 5, F_2 = 8, Q = 8$, and $P = 5$. Panels ($e$) and ($f$) demonstrate initial-condition-dependent convergence: trajectories evolve to strategy $N$ (non-adoption) and $C$ (complete adoption) under distinct initial frequencies, respectively, as predicted by the bistable regime in \ref{appendix2} (Case 8).}
\end{figure}
%==================================================================

%%example 4
\begin{example}
\begin{enumerate}
    \renewcommand{\labelenumi}{$(\theenumi)$}
    \item When $Q<M-R<P, M>N>R,$ and $P(M-N)+Q(N-R)>PQ$, as in the previous example, there are three corner equilibrium points and one equilibrium point on the $CN$ boundary on the phase plane, with no interior equilibrium point. The equilibrium point on the $CN$ boundary is stable, while the other three corner equilibrium points are unstable. For relevant theoretical proofs, please refer to Case 11 in \ref{appendix2}.
    \item Under $ Q < M - R < P \land M > R > N $, three unstable corner equilibrium points coexist with a stable boundary equilibrium on $CN$. Phase trajectories exhibit boundary-driven convergence (Fig. \ref{fig4}(b)). Theoretical foundations are detailed in \ref{appendix2} (Case 13).
\end{enumerate}
As shown in Figure \ref{fig4}, regardless of the specific scenario, farmers will coexist with a certain proportion of completely adopting and not adopting conservation tillage technology. The strategies of partially adopting conservation tillage technology will disappear.
\end{example}

%%example 5
\begin{example}
\begin{enumerate}
    \renewcommand{\labelenumi}{$(\theenumi)$}
    \item When $ P < M - R < Q \land M > N > R \land P(M - N) + Q(N - R) < PQ $, the system has five equilibria: three vertex points (only $(1,0,0)$ stable), one boundary saddle on $CN$, and an unstable interior equilibrium. Despite multistability, convergence to complete conservation tillage adoption occurs (Fig. \ref{fig4}(c)), as proven in \ref{appendix2} (Case 7).
    \item For \( Q < M - R < P \land M > N > R \land P(M - N) + Q(N - R) < PQ \), the unstable boundary equilibrium and corner equilibrium points contrast with a stable interior equilibrium point. This regime achieves three-strategy coexistence through asymptotically stable mixing ratios (Fig. \ref{fig4}(d)), as rigorously derived in \ref{appendix2} (Case 12).
\end{enumerate}\par
\end{example}

%======================= fig 4 ==========================
\begin{figure}[H]
\centering
\includegraphics[width=0.8\textwidth]{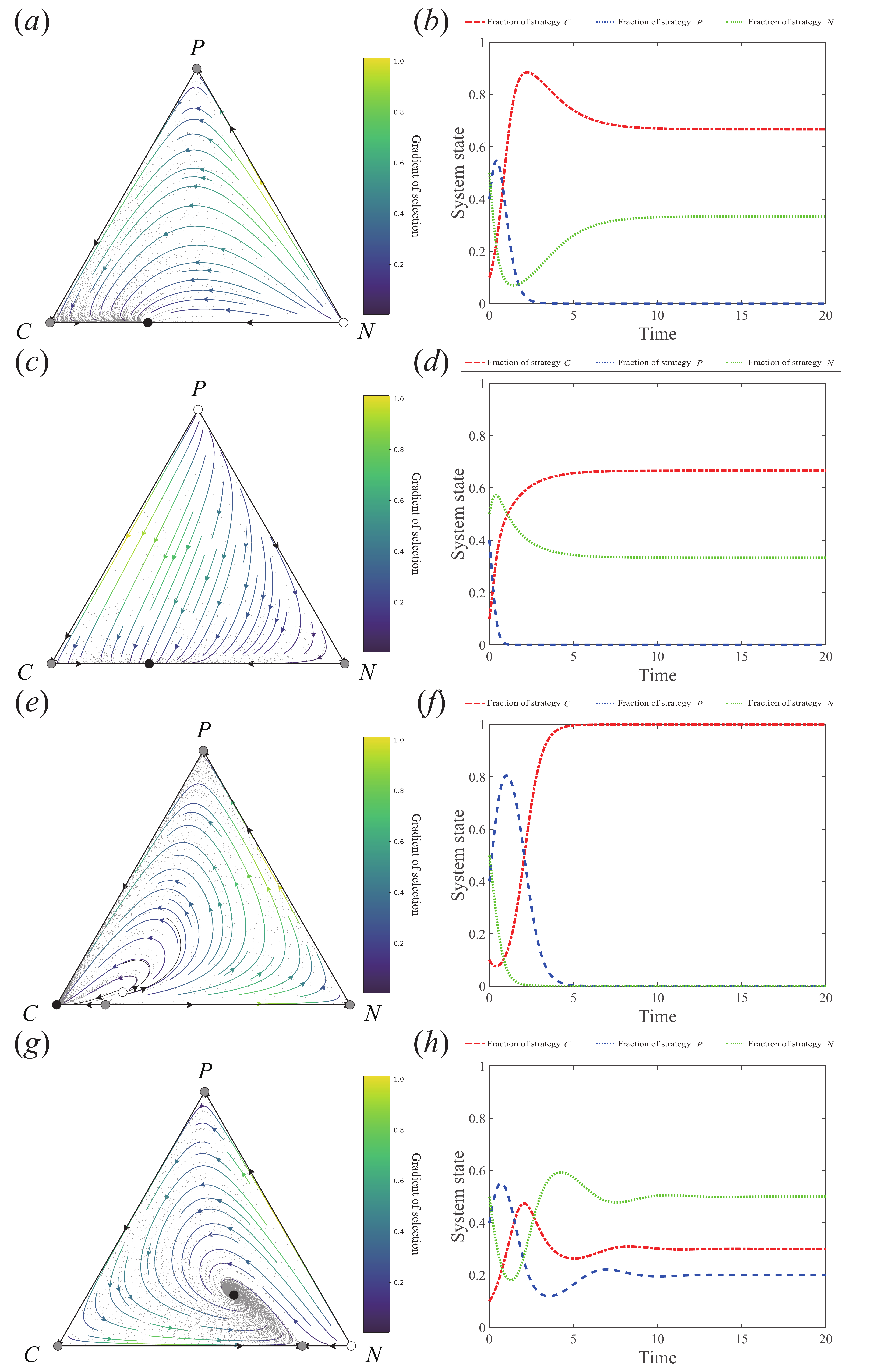}
\caption{\label{fig4} In Panel ($a$) and ($b$), $Q = 5, P=8, M=17$ and $N = 14$, in Panel ($c$) and ($d$), $Q = 5, P=8, M=17$ and $N = 8$, in Panel ($e$) and ($f$), $Q = 10, P=4, M=15$ and $N = 13$, and in Panel ($g$) and ($h$), $Q = 4, P=10, M=15$ and $N = 13$. Other parameter values include: $R = 10, C_2 = 5, F_1 = 5,$ and $F_2 = 8$.}
\end{figure}
%==================================================================

\section{Conclusion and Discussion}
\label{sec:Conclusion}
This study constructs an evolutionary game model based on evolutionary dynamics to analyze the strategic interactions among farmer groups during the adoption of CTT. The model incorporates three key dimensions: (1) time preference, quantified by the discount rate $r$, reflecting the extent to which farmers discount the long-term benefits of CTT; (2) the lemon market effect, characterized by lemon market benefit $P$ and loss $Q$, reflecting differences in the market competitiveness of agricultural products under different CTT adoption strategies; and (3) technological externalities, capturing the spillover effects of individual decisions on group benefits.

Research indicates that an individual farmer's degree of time preference constitutes a key intrinsic factor influencing their adoption decision regarding CTT. Specifically, farmers with a higher time preference tend to heavily discount the long-term ecological benefits and economic returns when evaluating CTT, leading to an underestimation of its future cumulative value. This results in significantly lower willingness to adopt and reduced adoption levels. This cognitive bias, operating within a discounting decision framework, seriously hinders the promotion of CTT. Consequently, when formulating technology promotion strategies, governments need to account for the heterogeneity in farmers' time preferences and enhance their understanding of the long-term benefits associated with CTT.

Secondly, the contrast relationship in the lemon market effect influences the stable state of equilibrium points within the population, determining the dynamic equilibrium of the system’s evolutionary path. When $P > Q$, the three strategies—complete adoption, partial adoption, and non-adoption—can coexist in stable proportions, reflecting healthy competition under symmetric market information. When $P < Q$, the three equilibrium points and the boundary form a heteroclinic cycle, and it becomes impossible to predict which strategy will dominate. The attractive heteroclinic cycle reveals a type of unpredictability inherent to deterministic systems, distinct from chaotic phenomena; in this case, breaking information asymmetry is necessary. It is noteworthy that when $P = Q$, the system is a conservative Hamiltonian system; the frequency of farmers choosing the three strategies exhibits periodic oscillations, and stable periodic solutions exist. This indicates that the adoption level of CTT will exhibit periodic oscillations fluctuating with market supply and demand but will not spontaneously converge to the ideal equilibrium. This implies that relying solely on market regulation cannot achieve the green transformation of agriculture, and government intervention possesses necessity.

The ideal goal for achieving green agricultural development is for all farmers to completely adopt CTT. However, attaining this goal requires satisfying specific algebraic conditions. Our derivation demonstrates that two conditions can each enable the system to reach the ideal state. Condition 1: When $M > N > R$ and $P(M - N) + Q(N - R) \neq PQ$, the system can achieve the ideal state. This means the government must ensure that the benchmark return for complete adoption is the highest, partial adoption is intermediate, and non-adoption is the lowest, thereby providing farmers with a step-by-step ladder for technological improvement. Simultaneously, satisfying $P(M - N) + Q(N - R) \neq PQ$ ensures the system avoids entering periodic oscillations, preventing farmers from repeatedly switching between different strategies. Condition 2: When $M - R > Q, M - R > P$, and $M > N$, the system can achieve the ideal state. This sets clear target thresholds for policy intervention: it requires that the net technology return ($M - R$) must cover the lemon market loss $Q$ caused by information
asymmetry, that the net technology return from complete adoption ($M - R$) outstrips the fraudulent lemon market benefit $P$ obtained by non-adopters, ensuring the non-adoption strategy cannot gain an advantage through fraudulent returns. Concurrently, the benchmark return for complete adoption $M$ must be higher than that for partial adoption $N$, preventing farmers from stagnating in the partial adoption state.

Based on the above analysis, the government should coordinate the following three measures to promote the adoption of CTT: First, break down agricultural product market information barriers to ensure premium pricing for quality products. Establishing a traceable eco-labeling certification system will enhance market transparency, increase consumers' willingness to pay for green agricultural products, and effectively elevate green price premiums. This creates market incentives for farmers adopting CTT. Second, provide systematic support to lower the barriers and risks of technology adoption. Implement targeted green technology subsidy policies to reduce farmers' CTT adoption costs. Develop specialized agricultural service systems to address farmers' technical application challenges and lower the operational barriers of CTT. Establish dedicated agricultural insurance products to mitigate risks such as yield fluctuations potentially encountered during the initial stages of CTT promotion, alleviating concerns among farmers with high time preference regarding short-term income uncertainty. Organize visits to demonstration bases and invite successful farmers to share their experiences, thereby enhancing farmers' objective understanding of the long-term benefits of CTT. Third, strengthen institutional constraints to guide the transformation of agricultural production practices. Impose environmental taxes on traditional farming practices that rely excessively on chemical fertilizers and pesticides, thereby internalizing their negative externalities and incentivizing farmers to adopt CTT. Explore the establishment of an ``Environmental Tax Modulation and Rebate Mechanism", where the revenue collected from these environmental taxes is specifically allocated to create a reward fund. This fund would provide rebates or rewards to farmers who comply with standards in implementing CTT, thereby creating a virtuous cycle of “polluters pay and conservators benefit.”

Applying game theory to solve practical problems in agricultural technology extension represents a significant research direction\citep{2020Evolution}. This study pioneers the introduction of evolutionary game theory into the domain of CTT adoption, thereby overcoming the limitations inherent in traditional static analytical frameworks. It offers valuable insights for promoting farmer adoption of CTT and advancing agricultural sustainability. However, it should be noted that this study has limitations, as we considered only three discrete adoption strategies. Future research could incorporate a continuous strategy space reflecting varying levels of CTT adoption. Analyzing its impact on system stability and convergence paths would advance our understanding of the complex systems driving agricultural green transition and ultimately provide more actionable scientific guidance for policymaking.

\appendix
\section{Boundary and internal equilibrium points}
\label{appendix1}
Assuming the existence of an equilibrium point on the $PN$ boundary, namely $x=0$, $f_y = f_z$, we can conclude that $N=R$.\par
Assuming the existence of an equilibrium point on the $PC$ boundary, namely $z=0$, $f_x = f_z$, we can conclude that $N=M$.\par
Assuming the existence of an equilibrium point on the $CN$ boundary, namely $y=0$, $f_x = f_y$, we can conclude that $x_0=\frac{M-R-Q}{P-Q}$ and $z_0=1-\frac{P-M+R}{P-Q}$. \par
Assuming the existence of an internal equilibrium point where $f_x = f_y = f_z$, we can conclude that 
$x^*=\frac{N-R}{P}$, $z^*=\frac{M-N}{Q}$ and $y^*=1-x^*-z^*$.

\section{Conditions for the stability of equilibrium points}
\label{appendix2}
This appendix provides the stability conditions of five equilibrium points under different parameter values. It includes a thorough analysis of thirteen different parameter combinations, and the stability conditions are presented in the form of a table.
\begin{table}[H]
    \centering 
    \caption{\label{tab1} Determinants and traces of Jacobian matrices at each fixed point}
    \begin{tabular}{ccc}
    \hline
    Point &  det$J$   &  tr$J$   \\
    \hline
    $(1,0,0)$       & $-(M-R-P)(N-M)$               & $-(M-R-P)+ N-M$ \\
    $(0,1,0)$       & $(M-N)(R-N)$                  & $M-2N+R$\\
    $(0,0,1)$       & $(M-R-Q)(N-R)$                & $M-2R-Q+N$\\
    $(x_0,0,z_0)$ & $[x_0(1-x_0)(Q-P)](N-R-x_0P)$   & $x_0(1-x_0)(Q-P)+N-R-x_0P$\\
    $(x^*,y^*,z^*)$ &$x^*y^*(1-x^*-y^*)PQ$          & $x^*(1-x^*-y^*)(Q-P)$\\
    \hline
    \end{tabular}
\end{table}

\section*{The theoretical analysis results for the 13 cases}
\begin{enumerate}
    \renewcommand{\labelenumi}{Case $\theenumi$.}
    % case 1
    \item If $M-R > Q, M-R > P$, and $M > N$, the evolutionary outcome of the system is
        \begin{table}[H]
        \centering 
        \caption{fixed points and their stability for the system in Case 1}
        \begin{tabular}{cccc}
        \hline
        Equilibrium point &  Criterion for judgment   & Stability\\
        \hline
        $(1,0,0)$       & $\lambda_1= -(M-R-P)<0, \lambda_2=N-M<0$ & ESS\\
        $(0,1,0)$       & $\lambda_1= M-N > 0$     & U\\
        $(0,0,1)$       & $\lambda_1= M-R-Q > 0$   & U\\
        $(x_0,0,z_0)$   &                          & None\\
        $(x^*,y^*,z^*)$ &                          & None\\
        \hline
        \end{tabular}
        \end{table}
        In the table, ESS represents the evolutionary stable strategy, $None$ indicates the absence of this point, and $U$ represents instability.
    % case 2
    \item If $M-R > Q, M-R > P$, $M<N$, and $R < N$, 
        \begin{table}[H]
        \centering 
        \caption{fixed points and their stability for the system in Case 2}
        \begin{tabular}{cccc}
        \hline
        Equilibrium point &  Criterion for judgment   & Stability\\
        \hline
        $(1,0,0)$       & $\lambda_2 > 0$      & U\\
        $(0,1,0)$       & $\lambda_1< 0,\lambda_2=R-N<0$  & ESS\\
        $(0,0,1)$       & $\lambda_1 > 0$   & U\\
        $(x_0,0,z_0)$   &                          & None\\
        $(x^*,y^*,z^*)$ &                          & None\\
        \hline
        \end{tabular}
        \end{table}
    % case 3
    \item If $M-R < Q, M-R < P$, $M<N$, and $R < N$, 
        \begin{table}[H]
        \centering 
        \caption{fixed points and their stability for the system in Case 3}
        \begin{tabular}{cccc}
        \hline
        Equilibrium point &  Criterion for judgment   & Stability\\
        \hline
        $(1,0,0)$       & $\lambda_2> 0$      & U\\
        $(0,1,0)$       & $\lambda_1< 0, \lambda_2<0$  & ESS\\
        $(0,0,1)$       & $\lambda_1 > 0$     & U\\
        $(x_0,0,z_0)$   &                          & None\\
        $(x^*,y^*,z^*)$ &                          & None\\
        \hline
        \end{tabular}
        \end{table}
    % case 4
    \item If $M-R < Q, M-R < P$, and $R > N$, 
        \begin{table}[H]
        \centering 
        \caption{fixed points and their stability for the system in Case 4}
        \begin{tabular}{cccc}
        \hline
        Equilibrium point &  Criterion for judgment   & Stability\\
        \hline
        $(1,0,0)$       & $\lambda_1 > 0$ & U\\
        $(0,1,0)$       & $\lambda_2 > 0$       & U\\
        $(0,0,1)$       & $\lambda_1 < 0, \lambda_2=N-R<0$ & ESS\\
        $(x_0,0,z_0)$   &                          & None\\
        $(x^*,y^*,z^*)$ &                          & None\\
        \hline
        \end{tabular}
        \end{table}
    % case 5
    \item If $M-R < Q, M-R < P$, and $M > N > R$, we have $P(M-N)+Q(N-R)<PQ$. Therefore, all three corner equilibrium points are unstable, and the stability of the interior equilibrium point depends on the magnitudes of P and Q.
        \begin{table}[H]
        \centering 
        \caption{fixed points and their stability for the system in Case 5}
        \begin{tabular}{cccc}
        \hline
        Equilibrium point &  Criterion for judgment   & Stability\\
        \hline
        $(1,0,0)$       & $\lambda_1 > 0$ & U\\
        $(0,1,0)$       & $\lambda_1> 0$      & U\\
        $(0,0,1)$       & $\lambda_2> 0$      & U\\
        $(x_0,0,z_0)$   &                           & None\\
        \hline
        \end{tabular}
        \end{table}
        When $P>Q$, we have det$J>0$ and tr$J<0$, therefore the equilibrium point is ESS. When $P<Q$, we have det$J>0$ and tr$J>0$, therefore the equilibrium point is unstable. When $P=Q$, we have det$J>0$ and tr$J=0$, therefore the equilibrium point is center.
    % case 6 
    \item If $P< M-R < Q, M > N > R$,  and $P(M-N)+Q(N-R)>PQ$, 
        \begin{table}[H]
        \centering 
        \caption{fixed points and their stability for the system in Case 6}
        \begin{tabular}{cccc}
        \hline
        Equilibrium point &  Criterion for judgment   & Stability\\
        \hline
        $(1,0,0)$       & $\lambda_1 <0, \lambda_2 <0$ & ESS\\
        $(0,1,0)$       & $\lambda_2 > 0$            & U\\
        $(0,0,1)$       & $\lambda_2 > 0$         & U\\
        $(x_0,0,z_0)$   & $\lambda_1=x_1(1-x_1)(Q-P)>0$  & U\\
        $(x^*,y^*,z^*)$ &                                & None\\
        \hline
        \end{tabular}
        \end{table}
    % case 7
    \item If $P< M-R < Q, M > N>R$, and $P(M-N)+Q(N-R)<PQ$, 
        \begin{table}[H]
        \centering 
        \caption{fixed points and their stability for the system in Case 7}
        \begin{tabular}{cccc}
        \hline
        Equilibrium point &  Criterion for judgment   & Stability\\
        \hline
        $(1,0,0)$       & $\lambda_1 > 0, \lambda_2 < 0$ & ESS\\
        $(0,1,0)$       & $\lambda_1 > 0$                & U\\
        $(0,0,1)$       & $\lambda_1 < 0, \lambda_2 > 0$ & U\\
        $(x_0,0,z_0)$   & $\lambda_1 > 0$                & U\\
        $(x^*,y^*,z^*)$ & det$J>0$ and tr$J>0$           & U\\
        \hline
        \end{tabular}
        \end{table}
    % case 8
    \item If $P< M-R < Q, M > N$, and $N < R$, we have $M>R>N$, 
        \begin{table}[H]
        \centering 
        \caption{fixed points and their stability for the system in Case 8}
        \begin{tabular}{cccc}
        \hline
        Equilibrium point &  Criterion for judgment   & Stability\\
        \hline
        $(1,0,0)$       & $\lambda_1 < 0, \lambda_2<0$ & ESS\\
        $(0,1,0)$       & $\lambda_2 > 0$            & U\\
        $(0,0,1)$       & $\lambda_1 < 0, \lambda_2 < 0$ & ESS\\
        $(x_0,0,z_0)$   & $\lambda_1>0$                  & U\\
        $(x^*,y^*,z^*)$ &                                & None\\
        \hline
        \end{tabular}
        \end{table}
    % case 9
    \item If $P< M-R < Q, M < N$, and $N > R$, we have $N>M>R$, 
        \begin{table}[H]
        \centering 
        \caption{fixed points and their stability for the system in Case 9}
        \begin{tabular}{cccc}
        \hline
        Equilibrium point &  Criterion for judgment   & Stability\\
        \hline
        $(1,0,0)$       & $\lambda_2 > 0$                  & U\\
        $(0,1,0)$       & $\lambda_1 < 0, \lambda_2 < 0$ & ESS\\
        $(0,0,1)$       & $\lambda_2 > 0$                & U\\
        $(x_0,0,z_0)$   & $\lambda_1>0$                  & U\\
        $(x^*,y^*,z^*)$ &                                & None\\
        \hline
        \end{tabular}
        \end{table}
    
    % case 10 Q< P
    \item If $Q< M-R < P$, and $N>M> R$, we have $P(M-N)+Q(N-R)<PQ$
        \begin{table}[H]
        \centering 
        \caption{fixed points and their stability for the system in Case 10}
        \begin{tabular}{cccc}
        \hline
        Equilibrium point &  Criterion for judgment   & Stability\\
        \hline
        $(1,0,0)$       & $\lambda_1 > 0$                  & U\\
        $(0,1,0)$       & $\lambda_1 < 0, \lambda_2 < 0$ & ESS\\
        $(0,0,1)$       & $\lambda_1 > 0$                & U\\
        $(x_0,0,z_0)$   & $\lambda_1 < 0, \lambda_2=N-R-x_1P>0$ & U\\
        $(x^*,y^*,z^*)$ &                                & None\\
        \hline
        \end{tabular}
        \end{table}
    % case 11
    \item If $Q< M-R < P, M>N>R$, and $P(M-N)+Q(N-R)>PQ$
        \begin{table}[H]
        \centering 
        \caption{fixed points and their stability for the system in Case 11}
        \begin{tabular}{cccc}
        \hline
        Equilibrium point &  Criterion for judgment   & Stability\\
        \hline
        $(1,0,0)$       & $\lambda_1 > 0$                & U\\
        $(0,1,0)$       & $\lambda_1 > 0$                & U\\
        $(0,0,1)$       & $\lambda_1 > 0$                & U\\
        $(x_0,0,z_0)$   & $\lambda_1 < 0, \lambda_2<0$ & ESS\\
        $(x^*,y^*,z^*)$ &                                & None\\
        \hline
        \end{tabular}
        \end{table}
    % case 12
    \item If $Q< M-R < P, M>N>R$, and $P(M-N)+Q(N-R)<PQ$
        \begin{table}[H]
        \centering 
        \caption{fixed points and their stability for the system in Case 12}
        \begin{tabular}{cccc}
        \hline
        Equilibrium point &  Criterion for judgment   & Stability\\
        \hline
        $(1,0,0)$       & $\lambda_1 > 0$                & U\\
        $(0,1,0)$       & $\lambda_1 > 0$                & U\\
        $(0,0,1)$       & $\lambda_1 > 0$                & U\\
        $(x_0,0,z_0)$   & $\lambda_1 < 0, \lambda_2>0$   & U\\
        $(x^*,y^*,z^*)$ & det$J>0$ and tr$J<0$           & ESS\\
        \hline
        \end{tabular}
        \end{table}
    % case 13
    \item If $Q< M-R < P$, and $M>R>N$, we have $P(M-N)+Q(N-R)>PQ$
        \begin{table}[H]
        \centering 
        \caption{fixed points and their stability for the system in Case 13}
        \begin{tabular}{cccc}
        \hline
        Equilibrium point &  Criterion for judgment   & Stability\\
        \hline
        $(1,0,0)$       & $\lambda_1 > 0$                & U\\
        $(0,1,0)$       & $\lambda_1 > 0$                & U\\
        $(0,0,1)$       & $\lambda_1 > 0$                & U\\
        $(x_0,0,z_0)$   & $\lambda_1 < 0, \lambda_2<0$ & ESS\\
        $(x^*,y^*,z^*)$ &                                & None\\
        \hline
        \end{tabular}
        \end{table}

\end{enumerate}
    
\section*{CRediT authorship contribution statement}
\textbf{Dingyi Wang:} Conceptualization, Methodology,Writing-original draft.\par
\textbf{Ruqiang Guo:} Conceptualization, Discussing, Writing reviewing and editing.\par
\textbf{Qian Lu:} Formal analysis, Writing - review \& editing, Supervision.\par

\section*{Declaration of Competing Interest}
The authors declare that they have no known competing financial interests or personal relationships that could have
appeared to influence the work reported in this paper.

\section*{Acknowledgment}
We would like to thank the referees for their careful reading and helpful comments. Our research was supported by the National Natural Science Foundation of China, under the project titled ``Research on the Demand Induction Mechanism for the Adoption of Conservation Tillage Technology: Organizational Support, Intertemporal Choices, and Incentive Effects” (Grant No. 71973105).

%%       \citet{<label>} ==> Jones et al. [21]
%%       \citep{<label>} ==> [21]
%%

\bibliographystyle{elsarticle-num-names} 
\bibliography{refs}




%%\bibitem[ ()]{}

%%\end{thebibliography}

\end{document}